\documentclass[aps,pra,preprintnumbers,amsmath,amssymb,superscriptaddress,floatfix,twocolumn]{revtex4-1}

\usepackage{graphicx}
\usepackage{indentfirst}
\usepackage{color,soul}
\usepackage{braket}
\usepackage{tabularx}

\begin{document}

\title{Enhanced Bell state measurement for efficient measurement-device-independent quantum key distribution using 3-dimensional quantum states}

\author{Yonggi Jo}\affiliation{Department of Physics, Sogang University, 35, Baekbeom-ro, Mapo-gu, Seoul 04107, Republic of Korea}
\affiliation{Research Institute for Basic Science, Sogang University, 35, Baekbeom-ro, Mapo-gu, Seoul 04107, Republic of Korea}
\author{Kwangil Bae}\affiliation{Department of Physics, Sogang University, 35, Baekbeom-ro, Mapo-gu, Seoul 04107, Republic of Korea}
\author{Wonmin Son}\affiliation{Department of Physics, Sogang University, 35, Baekbeom-ro, Mapo-gu, Seoul 04107, Republic of Korea}

\begin{abstract}
We propose an enhanced discrimination measurement for tripartite 3-dimensional entangled states in order to improve the discernible number of orthogonal entangled states. The scheme suggests 3-dimensional Bell state measurement by exploiting composite two 3-dimensional state measurement setups. The setup relies on state-of-the-art techniques, a multi-port interferometer and nondestructive photon number measurements that are used for the post-selection of suitable ensembles. With this scheme, the sifted signal rate of measurement-device-independent quantum key distribution using 3-dimensional quantum states is improved by up to a factor of three compared with that of the best existing setup.
\end{abstract}

\maketitle

\section{Introduction}

Quantum cryptography is a mature research field that exploits the principles of quantum mechanics to ensure its information theoretical security. The core protocol of quantum cryptography is quantum key distribution (QKD), which is the process of generating a secret key that is shared between two distant parties, called Alice and Bob. These parties are assumed to be exposed to a potential malicious eavesdropper, conventionally called Eve. Since the proposal of the first QKD protocols \cite{Bennett1984,Ekert1991}, many efforts have been made to improve the security of QKD based on quantum principles \cite{Deutsch1996,Mayer2001,Shor2000,Devetak2005}. Various types of QKD have also been experimentally demonstrated to date \cite{Breguet1994,Muller1995,Muller1997,Naik2000}.

The earliest proposal for QKD used 2-dimensional quantum states, called qubits \cite{Schumacher1995}. After this proposal, significant efforts were made to increase the key rate of QKD protocols. For example, protocols involving high-dimensional quantum states, called qudits, were introduced.  It is well known that higher-dimensional quantum states can carry more information per quantum. In fact, there have been many theoretical proposals of the exploitation of qudits in various types of quantum information processing, such as non-locality testing \cite{Collins2002,Son2004,Bae2016-1,Bae2016-2,Salavrakos2017} and quantum teleportation \cite{Braunstein2000,Son2001}. High-dimensional quantum states have been experimentally demonstrated in various quantum systems, energy-time eigenstates \cite{Thew2004,Khan2006}, multipath-entangled states \cite{Schaeff2012,Rossi2009,Lee2017,Wang2018}, and quantized orbital angular momentum (OAM) modes of photons \cite{Leach2002,Malik2016}.

Furthermore, applying high-dimensional states in QKD is known to increase the efficiency of key distribution under a potential attack by Eve in the ideal case \cite{Cerf2002,Durt2004,Ferenczi2012,Coles2016}. The results show that QKD based on qudits can achieve a higher key rate and a higher upper bound on the allowed error rate than the original QKD protocol can. Such protocols have been demonstrated using various photon degrees of freedom, such as energy-time states \cite{Pasquinucci2000,Khan2007,Mower2013,Nunn2013} and OAM modes \cite{Groblacher2006,Mirhosseini2015,Sit2017}.

Along another branch of investigation, the security of the original QKD system has been scrutinized in more detail. Device-independent QKD (DI-QKD) has been proposed to extend the notion of ultimate security in the device attack scenario \cite{Acin2006,Acin2007,Pironio2009,Hanggi2010,Masanes2011,Lim2013,Vazirani2014}. In this protocol, Alice and Bob can generate a secret key without any a priori assumptions regarding device performance. This scheme is designed to protect the key from side-channel attack when the measurement device is not very reliable. In this case, the security of the protocol is guaranteed only by nonlocal correlations as identified by the Clauser-Horne-Shimony-Holt (CHSH) inequality \cite{Clauser1969}. However, the DI-QKD protocol is not easy to be implemented in practice since it requires a loophole-free Bell test experiment, which poses high technological demands \cite{Giustina2013,Christensen2013,Larsson2014,Hensen2015,Rosenfeld2017}.

To compensate for this practical difficulty, a measurement-device-independent QKD (MDI-QKD) protocol was proposed in 2012 \cite{Lo2012}. This protocol can be more easily implemented than DI-QKD can because the MDI-QKD procedure does not rely on entanglement. In MDI-QKD, an untrusted third party, called Charlie, maintains quantum state detectors separately from Alice and Bob. After Alice and Bob send quantum states, {\it e.g.} single photon states, to Charlie, he performs a Bell state measurement (BSM) on the incoming photons to generate the correlation between Alice and Bob. In this protocol, Charlie acts as a referee to build up the necessary correlation. Due to the selective construction of the correlation, an eavesdropper who attacks the detector cannot obtain exact information about the secret key. For this reason, MDI-QKD possesses unbounded security against any detector attacks. Using MDI-QKD, most side-channel attacks made possible by detector imperfections can be resisted  \cite{Lo2014}.

However, MDI-QKD suffers from a low secret key rate. When the BSM in MDI-QKD is performed using linear optical elements, the success probability of the setup is only 50\% \cite{Lutkenhaus1998}. In the original QKD protocol, the secret key can be extracted when Alice and Bob use the same encoding basis, which is true for half of the generated ensembles on average. However, in MDI-QKD, Alice and Bob can share information only when they use the same encoding basis and the BSM is successful, meaning that 75\% of their trials must be discarded. This shortcoming makes this key distribution scheme quite inefficient.

To increase the key rate, aversion of MDI-QKD using high-dimensional quantum states was recently studied \cite{Jo2016,Hwang2016}, and a performance increase was demonstrated for MDI-QKD using 3-dimensional quantum states, called qutrits, instead of qubits \cite{Jo2016}. However, the practicality of this scheme is still questionable. This is because a generalized BSM of a bipartite high-dimensional maximally entangled state cannot be implemented using only linear optical elements \cite{Dusek2001,Calsamiglia2002}. For the implementation of high-dimensional entangled state discrimination, multi-mode quantum scissors \cite{Goyal2013} and a linear optical setup for multipartite high-dimensional entangled state discrimination \cite{Goyal2014} have been proposed, although their success probabilities are still not sufficient for the implementation of efficient high-dimensional MDI-QKD.

In the present work, we propose an efficient discrimination setup for tripartite entangled qutrit states to enhance the discernible number of orthogonal entangled states. 
The setup relies on state-of-the-art techniques, a multi-port interferometer called a tritter \cite{Reck1994}, and nondestructive photon number measurements \cite{imoto1987,Nogues1999}. The generation of multipartite high-dimensional entangled photonic states has not been well studied even now, and such states have been experimentally demonstrated only very recently \cite{Malik2016,Krenn2016,Krenn2017,Goyal2014}. With the technologies listed above, we construct a setup that can discriminate subsets of tripartite entangled qutrit states and show that qutrit MDI-QKD can be implemented using this setup. Moreover, we generalize the measurement to $d$-dimensional $d$-photon entangled state discrimination measurement and analyse the secret key rate of high-dimensional MDI-QKD using the generalized setup.

This article is organized as follows. We present a schematic description of the tripartite entangled qutrit state discrimination setup and the set of entangled states that the setup can discriminate. We also present schematic descriptions of the teleportation process and MDI-QKD using qutrit states with the setup we propose. We analyse the security of the MDI-QKD protocol using qutrit states as well. We generalize the proposed setup to a setup for $d$-dimensional $d$-photon entangled state discrimination and analyse the secret key rate of high-dimensional MDI-QKD. The efficiency of the proposed MDI-QKD setup when experimental factors are considered is described as well.

\section{Results}\label{SecResult}

\subsection{Tripartite entangled qutrit state discrimination setup}

\begin{figure}[!h]\centering{\includegraphics[width=0.43\textwidth]{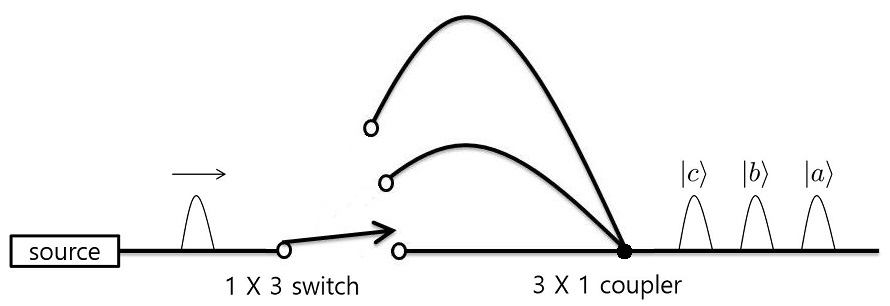}}
\caption{A schematic diagram of the time-bin qutrit generating source \cite{Pasquinucci2000}. The diagram is based on a figure from our previous work \cite{Jo2016}. In this setup, the photon injection time to the output port is controlled by three different delay lines and a photonic path switch. $\ket{a}$, $\ket{b}$, and $\ket{c}$ denote the time states when the photon passes through the shortest, intermediate, and longest delay lines, respectively.}
\label{fig3dsource}
\end{figure}

Before we introduce the discrimination setup, we describe the physical system we are considering and introduce the related notation. We consider 3-dimensional photonic states, where $\ket{a}$, $\ket{b}$, and $\ket{c}$ denote 3-dimensional orthonormal states. Such 3-dimensional quantum states can be realized by exploiting various degrees of freedom, such as, high-dimensional time-bin states \cite{Khan2006,Pasquinucci2000} and OAM modes of single photons \cite{Leach2002,Malik2016}. As an example, Fig.~\ref{fig3dsource} shows a schematic setup for generating 3-dimensional time-bin states \cite{Pasquinucci2000}. The source generates a single photon and the photon is injected into a delay line. The delay lines have different lengths, and the user can choose into which of the delay lines the photon is injected. $\ket{a}$, $\ket{b}$, and $\ket{c}$ denote the photonic time-bin states when the photon passes though the shortest, intermediate, and longest delay lines, respectively.

With regard to the qutrit states, we will focus on tripartite entangled states. To realize interference among three photons, we use a multi-port interferometer called a tritter \cite{Reck1994}. The tritter we consider has three input ports and three output ports. We consider the three-path operation described in Eq.~(\ref{unitary3}):
\begin{align}\label{unitary3}
\begin{pmatrix}
\hat{a}_{3}^{\dagger} \\
\hat{a}_{4}^{\dagger}  \\
\hat{a}_{5}^{\dagger}
\end{pmatrix}=\hat{U}_{3}\begin{pmatrix}
\hat{a}_{0}^{\dagger} \\
\hat{a}_{1}^{\dagger}  \\
\hat{a}_{2}^{\dagger}
\end{pmatrix}
=\frac{1}{\sqrt{3}}\begin{pmatrix}
1 & 1 & 1\\
1 & \omega & \omega^2 \\
1 & \omega^2 & \omega
\end{pmatrix}\begin{pmatrix}
\hat{a}_{0}^{\dagger} \\
\hat{a}_{1}^{\dagger}  \\
\hat{a}_{2}^{\dagger}
\end{pmatrix}
\end{align}
where $\hat{a}^{\dagger}_{y}$ is a photon creation operator on path $y$ and $\hat{U}_{3}$ is a three-dimensional discrete Fourier transformation defined in terms of $\omega$, which, in turn, is defined as $\omega=\exp(2 \pi i/3)$. The input and output ports are distinguished by the subscripted numbers $0$, $1$, and $2$ and $3$, $4$, and $5$, respectively.

In our notation, a single photon that is time-bin mode $x$ in the port labelled $y$ is represented by $\hat{a}_{xy}^{\dagger}\ket{0}=\ket{x_{y}}$. We consider nine states, as given in Eq.~(\ref{Estates}):
\begin{align}
&\ket{\Psi_{i}}=\frac{1}{\sqrt{6}}\sum_{j=0}^{2}\omega^{2ij}\ket{a_{j}}(\ket{b_{j+1},c_{j+2}}-\ket{b_{j+2},c_{j+1}})\nonumber\\\label{Estates}
&\ket{\Psi_{3+i}}=\frac{1}{\sqrt{6}}\sum_{j=0}^{2}\omega^{2ij}\ket{a_{j}}(\ket{b_{j},c_{j+1}}-\ket{b_{j+1},c_{j}})\\
&\ket{\Psi_{6+i}}=\frac{1}{\sqrt{6}}\sum_{j=0}^{2}\omega^{2ij}\ket{a_{j}}(\ket{b_{j+2},c_{j}}-\ket{b_{j},c_{j+2}}),\nonumber
\end{align}
where $i\in\{0,~1,~2\}$, $\omega$ is defined in Eq.~(\ref{unitary3}), and we omit $($mod $3)$ in the subscripts on the right-hand side. In the states described in Eq.~(\ref{Estates}), the three photons are in different time-bin modes, $\ket{a}$, $\ket{b}$, and $\ket{c}$. The tripartite entangled qutrit states $\ket{\Psi_{0}}$, $\ket{\Psi_{1}}$, and $\ket{\Psi_{2}}$ are the quantum states in which each photon exists in a separate input port, and the other states are those in which two photons of different time-bin modes exist in one port and the other photon is in another port. The orthogonality of the nine tripartite entangled qutrit states is easily provable.

\begin{figure}[!t]\centering{\includegraphics[width=0.43\textwidth]{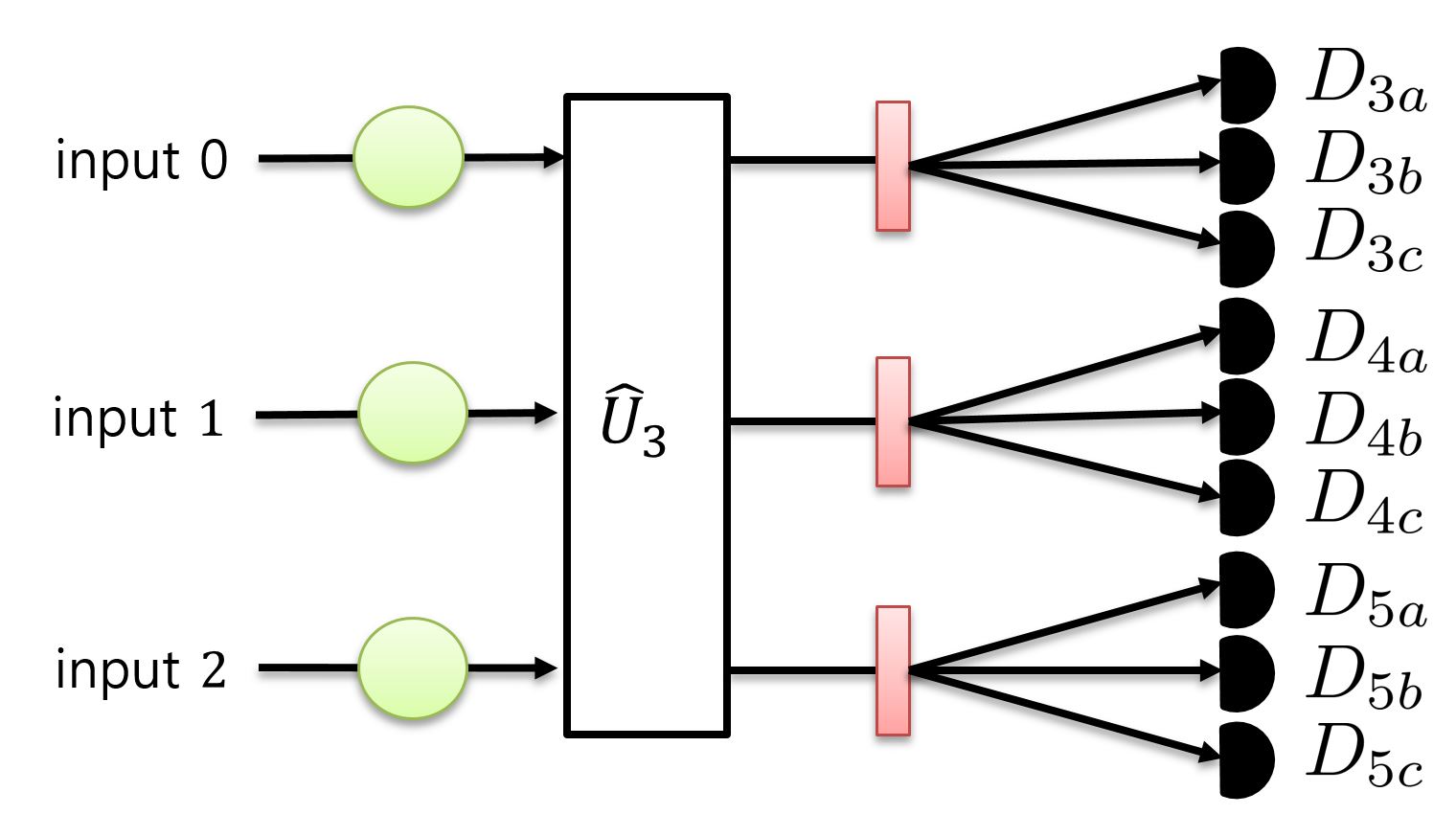}}
\caption{A schematic diagram of the measurement setup. $\hat{U}_{3}$ denotes the tritter setup described in the Methods section; the pink rectangular boxes represent state discrimination elements, which correspond to polarization beam splitters in polarization qubit experiments; the black semicircles denote on-off detectors; and the green circles on the input ports denote nondestructive measurement devices for detecting the parity of the photon number. $D_{xy}$ denotes a detector corresponding to a single photon state $\ket{y_{x}}$.}
\label{fig3dmeasurement}
\end{figure}

A schematic diagram of the entire tripartite entangled qutrit state discrimination setup is shown in Fig.~\ref{fig3dmeasurement}. Before photons that are in one of the states described in Eq.~(\ref{Estates}) are injected into the tritter, nondestructive photon number measurements are performed on each input port for post-selection. The details of the post-selection process will be described later. Subsequently, the photons enter the tritter, and the tritter performs the $\hat{U_{3}}$ operation on the photons. After interference, we perform a qutrit state discrimination measurement on each output port. Since we are using the time-bin modes of single photons, the states can be discriminated by measuring the arrival times of the photons at on-off detectors \cite{Marcikic2004,Donohue2013} or by using ultrafast optical switches \cite{Volz2012}. We obtain a certain combination of clicked detectors when one of our input states described in Eq.~(\ref{Estates}) is injected into the entire setup, yielding an output state as shown in Eq.~(\ref{Estatestrans}); the combination of clicked detectors for each state is given in Eq.~(\ref{detect}):
\begin{align}\label{detect}
\ket{\Psi_{3i}} \rightarrow\left\{\begin{matrix}
D_{3a},D_{4b},D_{5c}\\
D_{3a},D_{5b},D_{4c}\\
D_{4a},D_{3b},D_{5c}\\
D_{4a},D_{5b},D_{3c}\\
D_{5a},D_{3b},D_{4c}\\
D_{5a},D_{4b},D_{3c}
\end{matrix}\right. ,\\
\ket{\Psi_{3i+1}} \rightarrow\left\{\begin{matrix}
D_{3a},D_{3b},D_{4c}\\
D_{3a},D_{4b},D_{3c}\\
D_{4a},D_{4b},D_{5c}\\
D_{4a},D_{5b},D_{4c}\\
D_{5a},D_{3b},D_{5c}\\
D_{5a},D_{5b},D_{3c}
\end{matrix}\right. ,\nonumber\\
\ket{\Psi_{3i+2}} \rightarrow\left\{\begin{matrix}
D_{3a},D_{3b},D_{5c}\\
D_{3a},D_{5b},D_{3c}\\
D_{4a},D_{3b},D_{4c}\\
D_{4a},D_{4b},D_{3c}\\
D_{5a},D_{4b},D_{5c}\\
D_{5a},D_{5b},D_{4c}
\end{matrix},\right.\nonumber
\end{align}
where $i\in\{0,~1,~2\}$ and $D_{xy}$ denotes a click of the detector corresponding to the single-photon state $\ket{y_{x}}$. Each possible detection listed in Eq.(\ref{detect}) has an equal probability of 1/6. Since we are using on-off detectors, we cannot discriminate phase differences among the three states. This means that we can infer groups of entangled states from the measurement results, but we cannot discriminate exact entangled states. To discriminate the exact state, we post-select only those trials in which all three photons exist in different input ports by means of nondestructive photon number measurements. In a nondestructive measurement, the absorption of the photons during the measurement is ideally avoided, and the other degrees of freedom of the photons also remain unaffected. Nondestructive measurements of the photon number state can be successfully realized using nonlinear effects \cite{Sathyamoorthy2014,Xia2016} or an atom-cavity system \cite{Reiserer2013,OBrien2016}. The remaining states after post-selection are $\ket{\Psi_{0}}$, $\ket{\Psi_{1}}$, and $\ket{\Psi_{2}}$, which can be exactly discriminated from the combinations of clicked detectors given in Eq.~(\ref{detect}). The nondestructive photon number measurements do not affect other degrees of freedom of the photons in the case that all photons are distributed in different ports, so the interference pattern among the three photons in the tritter is the same when there is no nondestructive photon number measurement. We note that it is sufficient to use a nondestructive photon number parity measurement instead of a full nondestructive photon number measurement. If nondestructive photon number parity measurements on the input ports indicate an odd photon number, then it is guarantee that all of the photons are in different input ports.

\subsection{Path-encoded qutrit teleportation}

\begin{figure}[h!]\centering{\includegraphics[width=0.45\textwidth]{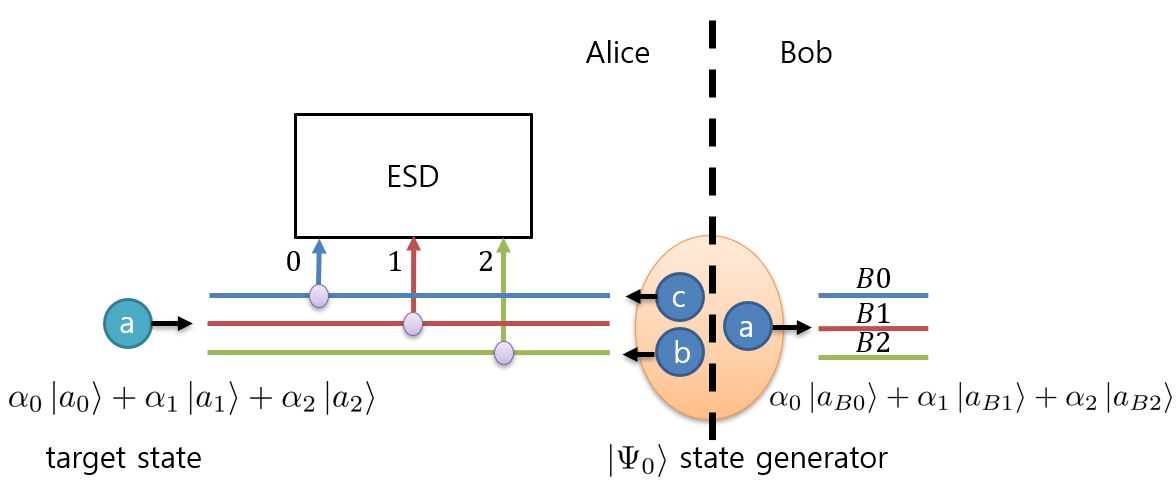}}
\caption{A schematic diagram of 3-dimensional path-encoded state teleportation. The purple circles represent $2\times 1$ couplers, fiber optical devices combining two paths into one path. ESD is the tripartite entangled qutrit state discrimination setup, and the numbers $0$,$1$, and $2$ denote the ports. The blue circles represent photons, and the letters on the circles denote the time-bin states of the photons. The numbers $0$, $1$, and $2$ correspond to the input ports connected to ESD, and $B0$, $B1$, and $B2$ denote the ports of the result states. Alice and Bob share the entangled state $\ket{\Psi_{0}}$ (orange circle), where Alice has the photons in time-bin states $b$ and $c$ and Bob has the photon in time-bin state $a$. Teleportation can be performed using the result of Alice's tripartite ESD measurement and Bob's corresponding unitary operation. The details of the scheme are described in the text.}\label{fig3dteleport}
\end{figure}

In this section, we show that a qutrit state teleportation protocol can be implemented with the proposed tripartite entangled qutrit state discrimination setup. The protocol is very similar to the previously studied teleportation protocol using linear optical elements \cite{Goyal2014}. Fig.~\ref{fig3dteleport} shows a schematic diagram of the qutrit teleportation protocol. The target state that we want to teleport is a path-encoded qutrit state, and its time-bin mode is $\ket{a}$. The target state can be written as shown in Eq.~(\ref{target}):
\begin{align}\label{target}
\alpha_{0}\ket{a_{0}}+\alpha_{1}\ket{a_{1}}+\alpha_{2}\ket{a_{2}}
\end{align}
where $\alpha_{i}$ is an arbitrary complex number satisfying the normalization condition, $\sum_{i=0}^{2}|\alpha_{i~}|^{2}=1$. One user, called Alice, possesses the target state. She wants to send that state to the other user, called Bob. To teleport the state, Alice and Bob share a photonic entangled state $\ket{\Psi_{0}}$. Previous studies have investigated the generation of multipartite photonic entangled states by means of an array of nonlinear crystals \cite{Krenn2017} and by means of tritter and nondestructive photon number measurements \cite{Goyal2014}. Alice has the target state and two photons generated from $\ket{\Psi_{0}}$ in time-bin states $\ket{b}$ and $\ket{c}$, and Bob has a photon generated from $\ket{\Psi_{0}}$ in time-bin state $\ket{a}$. Then, the entire system can be written as described in Eq.~(\ref{TeleTot}):

\begin{align}\label{TeleTot}
\ket{\Xi}=&(\alpha_{0}\ket{a_{0}}+\alpha_{1}\ket{a_{1}}+\alpha_{2}\ket{a_{2}})\\
&\otimes\left[\frac{1}{\sqrt{6}}\sum_{j=0}^{2}\ket{a_{Bj}}(\ket{b_{j+1},c_{j+2}}-\ket{b_{j+2},c_{j+1}})\right]\nonumber
\end{align}
where we omit (mod 3) in the subscripts. Alice applies the tripartite entangled qutrit state discrimination operation to her three photons. The entire system can be rewritten in the basis of Alice's measurement, as shown in Eq.~(\ref{TeleTotBasT}):

\begin{align}\label{TeleTotBasT}
\ket{\Xi}=&\frac{1}{3}\left[\sum_{i=0}^{2}\left(\ket{\Psi_{i}}\otimes\sum_{j=0}^{2}\alpha_{j}\omega^{ij}\ket{a_{Bj}}\right) \right. \\ &+\sum_{i=0}^{2}\left(\ket{\Psi_{3+i}}\otimes\sum_{j=0}^{2}\alpha_{j+1}\omega^{i(j+1)}\ket{a_{Bj}}\right) \nonumber\\ &\left. +\sum_{i=0}^{2}\left(\ket{\Psi_{6+i}}\otimes\sum_{j=0}^{2}\alpha_{j+2}\omega^{i(j+2)}\ket{a_{Bj}}\right)\right],\nonumber
\end{align}
where $\omega=\exp(2\pi i/3)$ and we omit (mod 3) in all subscripts except those of $\Psi$. After Alice's measurement, Alice sends the result of the measurement to Bob. Bob can recover the initial target state by performing a unitary operation on his state corresponding to Alice's result. The operation that Bob should perform for each possible result is specified in Table.~\ref{TbteleOP}. The operations that Bob needs to perform are two path rotating operations, which are described in Eq.~(\ref{teleBobop}):

\begin{align}\label{teleBobop}
\hat{P}_{1}&=\begin{pmatrix}
1 & 0 & 0\\
0 & \omega^{2} & 0 \\
0 & 0 & \omega
\end{pmatrix},\qquad
\hat{P}_{2}=\hat{P}_{1}^{~2}=\begin{pmatrix}
1 & 0 & 0\\
0 & \omega & 0 \\
0 & 0 & \omega^{2}
\end{pmatrix}.
\end{align}
The unitary operators $\hat{P}_{1}$ and $\hat{P}_{2}$ can be implemented with phase shifters on Bob's photonic paths $B0$, $B1$, and $B2$. For instance, to realize $\hat{P}_{1}$, the phase shifters should simultaneously apply no phase shift on $B0$, a phase shift of $\omega^{2}$ on $B1$, and a phase shift of $\omega$ on $B2$. After the unitary operation, Bob's photonic path state is as described in Eq.~(\ref{Bobresult}):
\begin{align}\label{Bobresult}
\alpha_{0}\ket{a_{B0}}+\alpha_{1}\ket{a_{B1}}+\alpha_{2}\ket{a_{B2}}.
\end{align}
This state is the same as the initial target state that is given in Eq.~(\ref{target}).

\begin{table}[t!]
\centering
\begin{tabularx}{\linewidth}{c|X X X c}
\hline
Alice's ESD result &\centering $\Psi_{0}$ &\centering $\Psi_{1}$ &\centering $\Psi_{2}$& \\
\hline
Bob's operation &\centering - &\centering $\hat{P}_{1}$ &\centering $\hat{P}_{2}$& \\
\hline
\end{tabularx}
\caption{Bob's unitary operations for qutrit teleportation corresponding to the result of Alice's entangled state discrimination measurement. The definition of each unitary operator is given in Eq.~(\ref{teleBobop}).}\label{TbteleOP}
\end{table}

\subsection{Measurement-device-independent quantum key distribution using qutrits}

In this section, we study the conceptual implementation of MDI-QKD using qutrit states. MDI-QKD is proposed to prevent side-channel attacks against imperfect measurement devices \cite{Lo2012}. In MDI-QKD, Alice and Bob, who want to share a secret key, send their encoded photonic states to an untrusted third party, Charlie. Charlie then performs a BSM to measure the correlation of the photons and announces the result. Alice (Bob) can infer the state that Bob (Alice) sent to Charlie from the announced result and her (his) own encoded state. Since the measurement setup detects only the correlation of the photons, an eavesdropper Eve cannot obtain the information that Alice and Bob share by attacking the measurement setup. It has been proven that QKD with high-dimensional quantum states has a higher secret key rate than that of traditional qubit QKD \cite{Cerf2002,Durt2004,Ferenczi2012,Coles2016}, and there are also studies that have shown that MDI-QKD with high-dimensional states has some advantages compared with qubit MDI-QKD \cite{Jo2016,Hwang2016}. The procedure for MDI-QKD using qutrit states was introduced in our previous work \cite{Jo2016}. Since the proposed setup is designed to discriminate tripartite entangled qutrit states, the quantum states in which Alice and Bob encode their information must be different from those used in the original protocol in order to exploit the setup. Here, we will show that MDI-QKD can be implemented with our proposed setup.

\begin{figure}[t!]\centering{\includegraphics[width=0.45\textwidth]{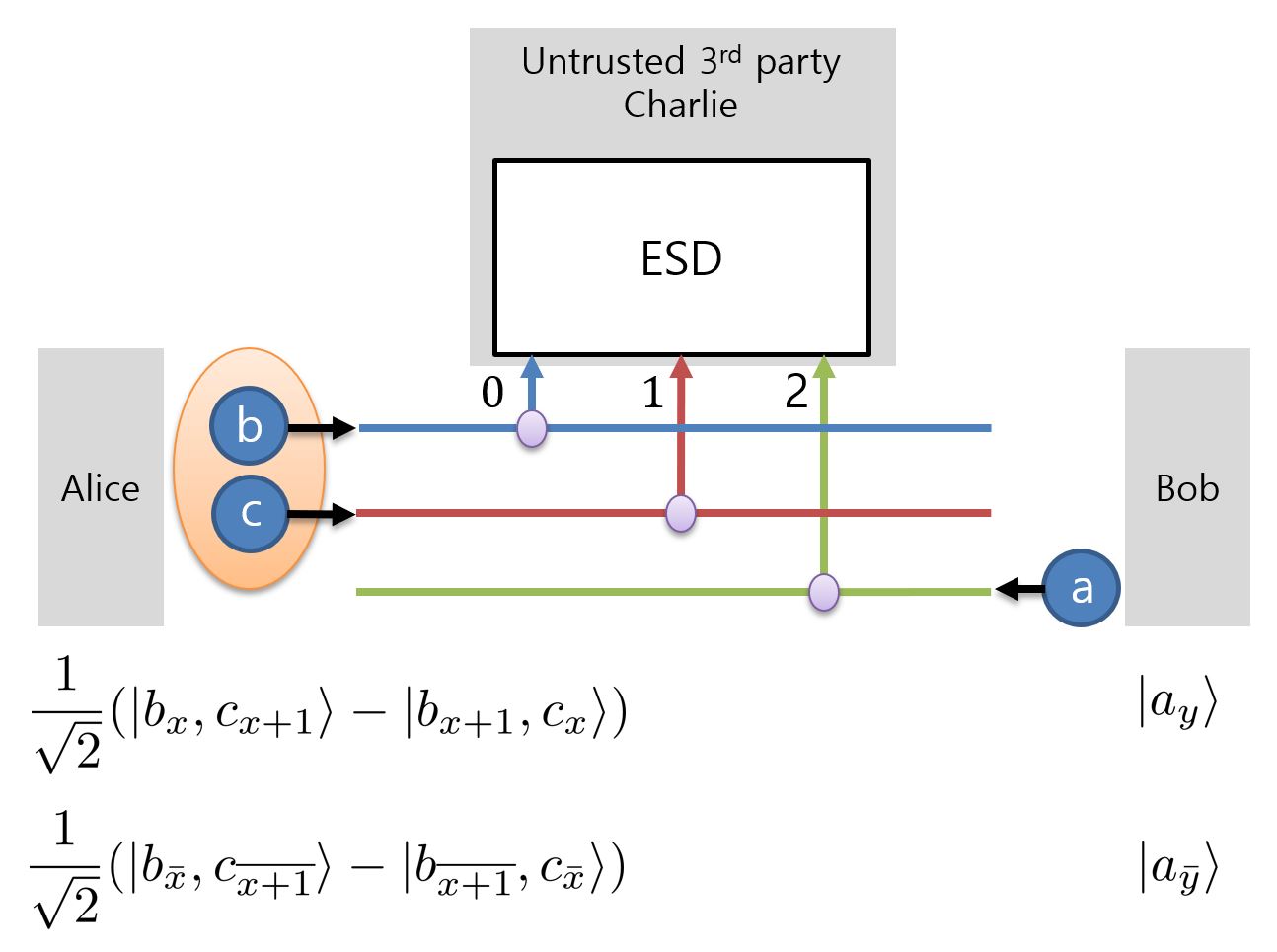}}
\caption{A schematic diagram of MDI-QKD using qutrits. The purple circles represent $2\times 1$ couplers, ESD denotes the entengled state discrimination measurement setup, and the numbers $0,1$, and $2$ denote three paths. Subscripts $x$ and $y$ in the states of Alice and Bob represent the labels of the paths on which the photons exist, meaning that $x$, $y\in \{0,1,2\}$ and we omit (mod $3$) in the subscripts. The photons in the large orange circle are in an entangled state. Alice generates a bipartite path-entangled photon pair in time-bin states $b$ and $c$ on paths $x$ and $x~+~1$ and sends them to an untrusted third party, Charlie. Bob generates a photon in state $a$ on path $y$ and also sends it to Charlie.The details of the states are described in the text.}\label{fig3dMDIQKD}
\end{figure}

In the original MDI-QKD protocol, Alice and Bob use two encoding bases at each site. As an extension of that protocol, our protocol also uses two encoding bases, a path-encoding basis and one of its mutually unbiased bases (MUBs). The condition for the two bases to be MUBs in three-dimensional Hilbert space, $|\braket{\bar{i}|j}|^{2}~=~1/3$, should be satisfied for all $i,j\in\{0,~1,~2\}$, where $\{\ket{0},~\ket{1},~\ket{2}\}$ and $\{\ket{\bar{0}},~\ket{\bar{1}},~\ket{\bar{2}}\}$ are orthonormal bases. 
In our physical system, $\{\ket{x_{0}},~\ket{x_{1}},~\ket{x_{2}}\}$ is a path-encoding basis, and $\{\ket{x_{\bar{0}}},~\ket{x_{\bar{1}}},~\ket{x_{\bar{2}}}\}$ is one of its MUBs, as defined in Eq.~(\ref{MUBrel}):

\begin{align}
\ket{x_{\bar{0}}}&=\frac{1}{\sqrt{3}}(\ket{x_{0}}+\ket{x_{1}}+\ket{x_{2}})\nonumber\\\label{MUBrel}
\ket{x_{\bar{1}}}&=\frac{1}{\sqrt{3}}(\ket{x_{0}}+\omega\ket{x_{1}}+\omega^{2}\ket{x_{2}})\\
\ket{x_{\bar{2}}}&=\frac{1}{\sqrt{3}}(\ket{x_{0}}+\omega^{2}\ket{x_{1}}+\omega\ket{x_{2}}).\nonumber
\end{align}
The states in the MUB can be generated with the tritter operation $\hat{U}_{3}$. For example, state $\ket{x_{\bar{0}}}$ can be generated by inputting the photonic state $\ket{x}$ into input port $0$ of the tritter depicted in the methods section.

A description of MDI-QKD using qutrit states is given as follows. Since the measurement setup projects an incoming state onto tripartite entangled states, the total number of encoded photonic states that Alice and Bob send to Charlie should be three. We assume that Alice sends two photonic states and that Bob sends one photonic state as shown in Fig.~\ref{fig3dMDIQKD}. First, Alice randomly chooses encoding information from among the ordinary numbers $1$, $2$ and $3$ and the bar numbers, $\bar{0}$, $\bar{1}$ and $\bar{2}$. After that, Alice generates the corresponding bipartite path entangled state. If Alice chooses a number $x$, Alice generates the state $(1/\sqrt{2})\left(\ket{b_{x},~c_{x+1}}-\ket{b_{x+1},~c_{x}}\right)$. Bob chooses a number $y$ from among the numbers Alice used and generates the state $\ket{a_{y}}$. Alice and Bob send their states to Charlie; then, Charlie performs tripartite entangled qutrit state discrimination on the incoming photons and announces the result through a public channel. Subsequently, Alice and Bob discard trials in which their encoding bases were different after a basis comparison through a public channel. The remaining data become the sifted key, and Alice and Bob can synchronize their information by performing the appropriate post-processing as described in our previous work \cite{Jo2016}.

To evaluate the usefulness of the proposed protocol, a security analysis of this protocol is necessary. Such an analysis can be performed through an inspection of the equivalent protocol using the entanglement distillation process (EDP) \cite{Deutsch1996,Shor2000,Devetak2005}. If the two parties, Alice and Bob, share the maximally entangled state generated via the EDP, then an eavesdropper cannot establish correlations between her state and the states of Alice and Bob \cite{Coffman2000}. In this sense, if Alice and Bob share a maximally entangled state, then they are assured that their protocol is secure against eavesdroppers. Thus, the security of the proposed protocol can be analysed with respect to the number of maximally entangled states generated via the EDP.

\begin{figure*}[t!]\centering{\includegraphics[width=0.7\textwidth]{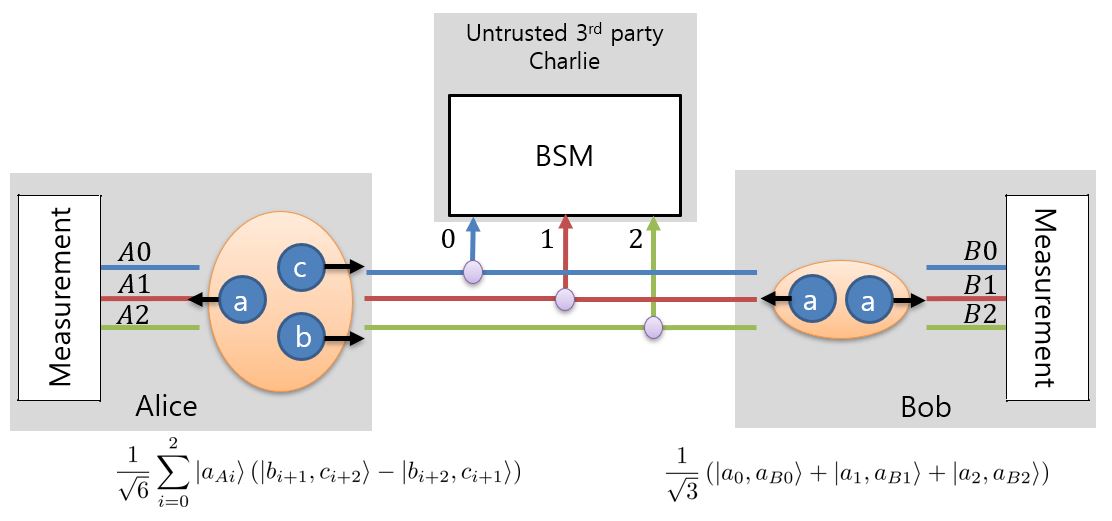}}
\caption{A schematic diagram of MDI-QKD using entangled qutrit states for security analysis. Alice prepares tripartite entangled qutrits and sends two of them, in states $b$ and $c$, to Charlie using state discrimination elements. Bob prepares bipartite maximally path-entangled qutrits with both photons in state $a$ and sends one of them to Charlie. Charlie performs tripartite entangled qutrit state discrimination (ESD) on the incoming photons and announces the result. By means of a unitary operation applied to Bob's state, Alice and Bob can share bipartite maximally entangled qutrits.}
\label{fig3dMDIQKDSP}
\end{figure*}

A schematic diagram of the equivalent protocol that exploits the maximally entangled state is shown in Fig.~\ref{fig3dMDIQKDSP}. In this protocol, we assume that Alice has a tripartite qutrit path-entangled state generator and that Bob has a bipartite qutrit path entangled state generator. Implementations of multipartite high-dimensional path-entangled states using linear optical elements and nondestructive photon number parity measurements \cite{Goyal2014} and using overlapping paths of photon pairs created in different crystals \cite{Krenn2017} have recently been proposed. The generation of bipartite high-dimensional path-entangled states using nonlinear crystals has previously been studied and demonstrated \cite{Rossi2009,Lee2017}. With regard to the states, Alice keeps the state that is in time-bin mode $a$ and sends the other photons to Charlie by using state discrimination elements, and Bob also keep one photon and sends the other to Charlie. Then the whole system can be described as written in Eq.~(\ref{QKDSPtotal}):

\begin{align}\label{QKDSPtotal}
\ket{\Xi}=&\frac{1}{\sqrt{6}}\sum_{i=0}^{2}\ket{a_{Ai}}(\ket{b_{i+1},c_{i+2}}-\ket{b_{i+2},c_{i+1}})\\
&\otimes \frac{1}{\sqrt{3}}\sum_{j=0}^{2}\ket{a_{j},a_{Bj}}.\nonumber
\end{align}
The entire system can be rewritten as given in Eq.~(\ref{QKDSPmtotal}):

\begin{align}\label{QKDSPmtotal}
\ket{\Xi}=&\frac{1}{3}\left\{\sum_{i=0}^{2}\left[\ket{\Psi_{i}}\otimes\frac{1}{\sqrt{3}}\sum_{j=0}^{2}\omega^{ij}\ket{a_{Aj},a_{Bj}}\right]\right.\\
&+\sum_{i=0}^{2}\left[\ket{\Psi_{i+3}}\otimes\frac{1}{\sqrt{3}}\sum_{j=0}^{2}\omega^{i(j+1)}\ket{a_{Aj},a_{B(j+1)}}\right]\nonumber\\\nonumber
&\left.+\sum_{i=0}^{2}\left[\ket{\Psi_{i+6}}\otimes\frac{1}{\sqrt{3}}\sum_{j=0}^{2}\omega^{i(j+2)}\ket{a_{Aj},a_{B(j+2)}}\right]\right\}.
\end{align}
In Eq.~(\ref{QKDSPmtotal}), the state $\ket{\Psi}$ represents the tripartite state that Alice and Bob sent to Charlie for the entangled state discrimination measurement. The bipartite states in the square brackets denote the path-entangled qutrit state that Alice and Bob share. With the proposed setup, the states $\ket{\Psi_{0}}$, $\ket{\Psi_{1}}$, and $\ket{\Psi_{2}}$ can be exactly discriminated. The other states cannot be exactly discriminated, so Alice and Bob should discard any trials in which Charlie's announced result is not one of $\ket{\Psi_{0}}$, $\ket{\Psi_{1}}$, and $\ket{\Psi_{2}}$. After sifting, Bob performs a unitary operation on his state based on Charlie's result; then, the final state Alice and Bob share is the maximally entangled state $1/\sqrt{3}\sum_{j=0}^{2}\ket{a_{Aj},a_{Bj}}$ when there is no error. Alice and Bob privately choose their measurement bases from between the photon path measurement and the corresponding MUB measurement and perform their measurements. The measurement results of Alice and Bob are strongly correlated only when Alice and Bob choose the same basis, so they discard any trial in which different bases are used after a basis comparison step conducted through public communication. The remaining data are correlated, so these data can be used as a secret key. Then, the security analysis of the MDI-QKD protocol using entangled qutrit states becomes equivalent to that of the QKD protocol using 3-dimensional maximally entangled states, which has already been studied \cite{Durt2004,Ferenczi2012,Jo2016}. According to the results, the secret key rate per sifted signal of the protocol, $r_{3}$,can be evaluated as shown in Eq.~(\ref{3dSK}):

\begin{align}\label{3dSK}
r_{3}\geq \log_{2}3-2Q-2H(Q)
\end{align}
where $H(x)$ is the Shannon entropy, defined as $H(x)=-x\log_{2}~x~-~(1~-~x)\log_{2}~(1~-~x)$, and $Q$ denotes the state error rate in the path-encoding basis. Each sifted signal is the result of a trial in which Alice and Bob generated the $\ket{\Psi_{0}}$ state and chose the same measurement basis. The error rate can be obtained as follows (number of signals that contain errors)/(number of sifted signals). An error corresponds to a case in which Alice and Bob share $\ket{\Psi_{x}}$ such that $x\neq 0$ at the end of the protocol.

\subsection{$d$-dimensional $d$-photon state discrimination setup and its efficiency}

\begin{figure}[h!]\centering{\includegraphics[width=0.45\textwidth]{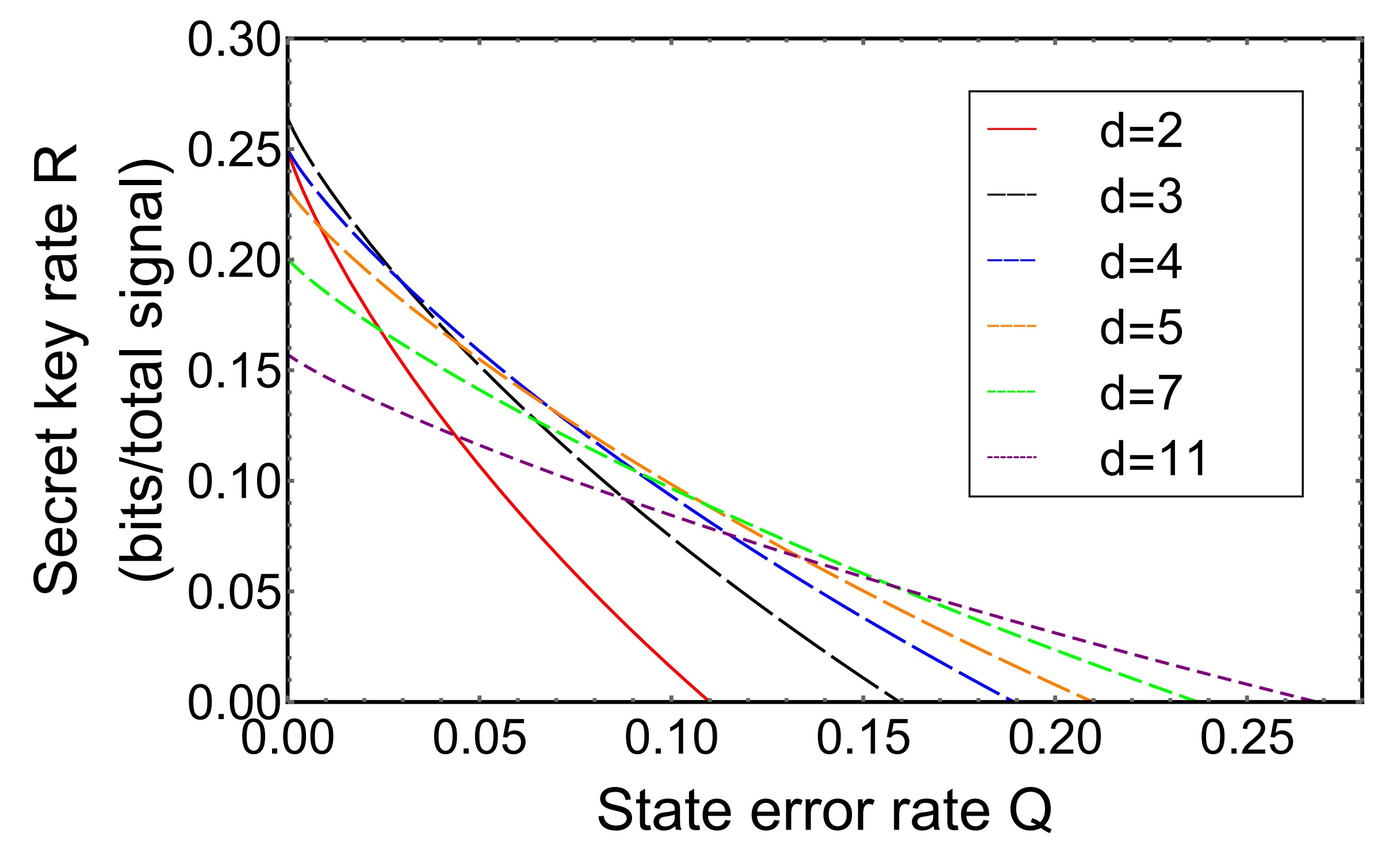}}
\caption{The secret key rate per total signal of $d$-dimensional MDI-QKD in the ideal situation. Since the number of states that can be discriminated via generalized $d$-dimensional $d$-photon entangled state discrimination is $d$ among the $d^{2}$ possible states considered, the success probability of entangled state discrimination is $1/d$ if all experimental factors are ignored. The secret key rate per total signal, $R$, can be obtained as follows: (secret key rate per sifted signal, $r$)$\times$(probability that a trial produces a sifted signal). The plot shows that only MDI-QKD using 3-dimensional quantum states has a higher secret key rate than that of the original MDI-QKD protocol at $Q=0$, the zero-error case.}
\label{figHDMDIKR}
\end{figure}

We investigate applications of a $d$-dimensional generalized setup and their efficiency. With the generalized setup and states, the qutrit teleportation scheme and qutrit MDI-QKD scheme described in previous sections can be extended to the teleportation of arbitrary $d$-dimensional path-encoded states and $d$-dimensional MDI-QKD, respectively. For instance, the steps of the $d$-dimensional MDI-QKD protocol are very similar to those of the qutrit MDI-QKD protocol, except that Alice and Bob should prepare $d$-dimensional information instead of 3-dimensional information and Alice and Bob use the states $\ket{A_{i}}$ and $\ket{0_{j}}$ to encode their information, where $\ket{A_{i}}$ is defined in the Methods section and $i,j\in\{0,~1,~2,~...,~d~-~1\}$. The security proof for $d$-dimensional QKD using maximally entangled states has already been studied \cite{Durt2004,Ferenczi2012}. Using the results, the secret key rate per sifted signal of $d$-dimensional QKD can be written as given in Eq.~(\ref{dkeyrate}):

\begin{align}\label{dkeyrate}
r_{d}\geq ~\log_{2}d+2(1-Q)\log_{2}(1-Q)+2Q\log_{2}\left(\frac{Q}{d-1}\right).
\end{align}
where $Q$ is the state error rate. As previously mentioned, MDI-QKD has the disadvantage of a lower secret key rate than that of the original BB84 protocol since Alice and Bob must discard more trials in MDI-QKD than in BB84 because of the success probability of the BSM. Here, we investigate the secret key rate per total signal of $d$-dimensional MDI-QKD. The secret key rate per total signal $R$ is obtained from (sifted signal rate)$\times$(secret key rate per sifted signal), where the sifted signal rate contains the probability that the $d$-dimensional $d$-photon entangled state discrimination succeeds and that Alice and Bob chose the same basis. In $d$-dimensional MDI-QKD, the number of possible combinations that Charlie can receive is $d^{2}$ since Alice and Bob each use $d$ different orthonormal states. Among the possible combinations, only the states in which all photons exist in different ports are post-selected by means of nondestructive photon number measurements. They are projected onto $\{\ket{\Phi_{i}}\bra{\Phi_{i}}|i~=~0,~1,~2,~...,~d-1\}$, and all other states are discarded. Thus, the success probability of $d$-dimensional $d$-photon entangled state discrimination is $1/d$ in the ideal situation in which we ignore all experimental factors and there is no eavesdropper. With this probability, the secret key rate per total signal of $d$-dimensional MDI-QKD can be calculated as $r_{d}/(2d)$, where the 2 in the denominator comes from the probability that Alice and Bob choose the same basis. The secret key rates per total signal of $d$-dimensional MDI-QKD with various values of $d$ are shown in Fig.~\ref{figHDMDIKR}. It is already known that high-dimensional QKD is more robust against state error than qubit QKD is \cite{Cerf2002,Durt2004,Ferenczi2012,Coles2016}, this phenomenon is also reflected in this plot. In the case of no error, when $Q=0$, only MDI-QKD using qutrit states has a higher key rate than that of the original qubit MDI-QKD protocol. The expression for the secret key rate per total signal when $Q=0$ is $(\log_{2}d)/d$. In this expression, the denominator increases linearly with the number of dimensions while the numerator increases logarithmically, so the secret key rate decreases in the high-dimensional case. From the plot, we can identify that qutrit MDI-QKD has the highest secret key rate per total signal when $0\leq Q\leq 0.0294$. This means that qutrits are best high-dimensional quantum states for MDI-QKD in the low-error range.

In real experiments, our proposed setup can fail even if the input state is one of $\{\ket{\Phi_{i}}|i~=~1,~2,~...,~d~-~1\}$ since the setup involves nondestructive photon number measurements and the success probability of such measurements is not 100\%. The generalized $d$-dimensional $d$-photon entangled state discrimination setup involves $d$ nondestructive photon number measurements, so the efficiency of this setup is exponentially affected by the success probability of these measurements. To investigate its usefulness, we compare the generalized setup with a ``Bell filter" that consists only of linear optical elements \cite{Goyal2014}. The Bell filter can discriminate $\ket{\Psi_{0}}$ regardless of dimensionality, so the sifted signal rate of $d$-dimensional MDI-QKD using the Bell filter is always $1/d^{2}$, where we ignore the probability that Alice and Bob choose the same basis since it is the same for all considered protocols. For the generalized setup we propose, the sifted signal rate is $1/d\times(\eta)^{d}$, where $\eta$ is the success probability of each nondestructive photon number measurement. For the sifted signal rates of the two setups to be the same, $\eta$ should be $\eta=(1/d)^{(1/d)}$; examples of the value of $\eta$ are 0.693, 0.707, and 0.725 for 3, 4, and 5 dimensions, respectively. Since the efficiency of nondestructive photon number measurements using an atom-cavity system was reported to be 0.66 in 2013 \cite{Reiserer2013}, the success probability of the proposed setup is lower than that of a Bell filter with current technology, even though the proposed setup can discriminate more states than the Bell filter can.

\section{Discussion}

We investigated a tripartite entangled qutrit state discrimination setup and its applications, especially teleportation and MDI-QKD. We showed that the proposed setup can discriminate three tripartite entangled qutrit states. We showed that the setup can be generalized to $d$-dimensional $d$-photon entangled state discrimination and that the secret key rate per total signal of $d$-dimensional MDI-QKD using the generalized setup is highest when qutrit states are used. We considered the efficiency of the nondestructive photon number measurements performed in the setup and calculated the efficiency bound of the proposed setup for which the sifted signal rate becomes higher than that of a Bell filter that consists of linear optical elements \cite{Goyal2014}. Since the success probability of the nondestructive photon number measurements using an atom-cavity system is 66\% \cite{Reiserer2013}, the proposed setup is not as efficient as a Bell filter with current technology. However, we expect that the proposed setup will be more efficient with future technologies.

\section{Methods}

\subsection{Tritter operation in the proposed setup}

\begin{figure}[!h]\centering{\includegraphics[width=0.45\textwidth]{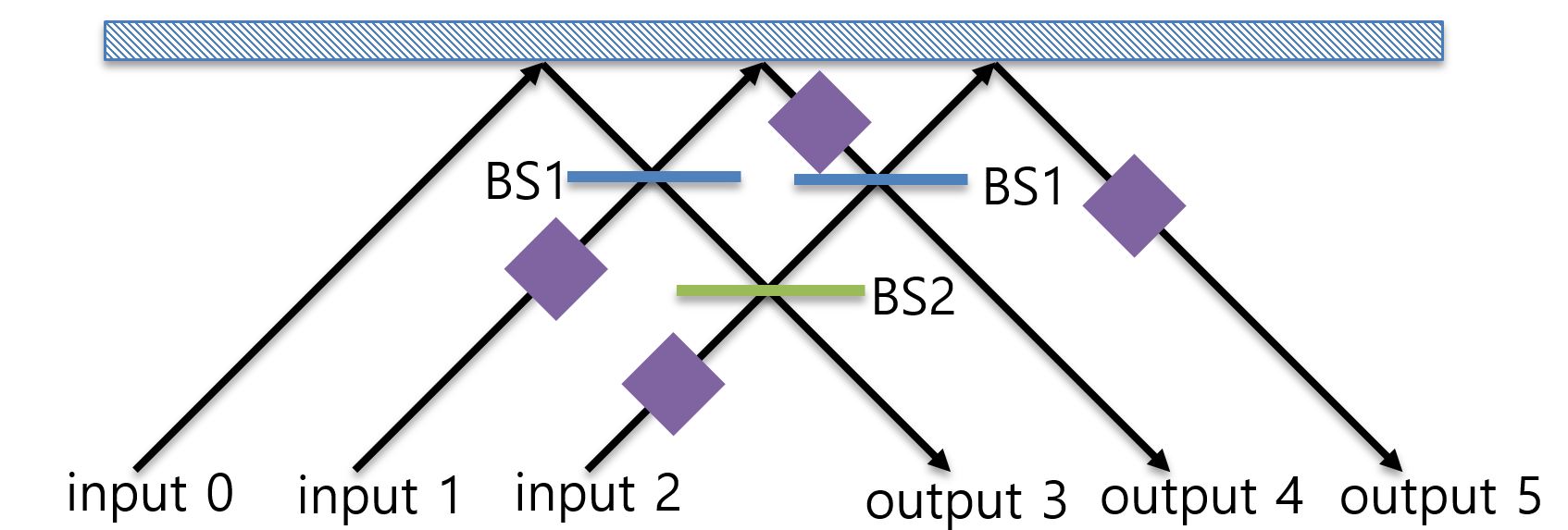}}
\caption{A schematic diagram of the tritter setup for realizing the unitary operator $\hat{U}_{3}$ described in Eq.~(\ref{unitary3}). The blue box represents a mirror, each BS1 (blue line) represents a 50:50 beam splitter, BS2 (green line) represents a beam splitter whose reflectivity is $1/3$, and the purple squares represent phase shifters.}
\label{fig3dtritter}
\end{figure}

A schematic diagram of the tritter that corresponds to the unitary operator $\hat{U}_{3}$ is shown in Fig.~\ref{fig3dtritter}. The tritter contains only linear optical elements: two 50:50 beam splitters, one beam splitter whose reflectivity is $1/3$ and several phase shifters. When the input states is one of the state described in Eq.~(\ref{Estates}), the output state after the tritter operation can be easily obtained from the input-output relation of the tritter given in Eq.~(\ref{unitary3}); the possible states are given in Eq.~(\ref{Estatestrans}):

\begin{align}
&\hat{U}_{3}\ket{\Psi_{3i}}\sim\frac{1}{\sqrt{6}}\sum_{j=3}^{5}\omega^{ij}\ket{a_{j}}(\ket{b_{j+1},c_{j+2}}-\ket{b_{j+2},c_{j+1}})\nonumber\\\label{Estatestrans}
&\hat{U}_{3}\ket{\Psi_{3i+1}}\sim\frac{1}{\sqrt{6}}\sum_{j=3}^{5}\omega^{ij}\ket{a_{j}}(\ket{b_{j},c_{j+1}}-\ket{b_{j+1},c_{j}})\\
&\hat{U}_{3}\ket{\Psi_{3i+2}}\sim\frac{1}{\sqrt{6}}\sum_{j=3}^{5}\omega^{ij}\ket{a_{j}}(\ket{b_{j+2},c_{j}}-\ket{b_{j},c_{j+2}}),\nonumber
\end{align}
where $i~\in~\{0,~1,~2\}$, the value of a subscript $j$ in Eq.~(\ref{Estatestrans}) is equal to as $3~+~[~j$ $($mod $3)]$ if a number in the subscript is larger than 5, and an unimportant global phase is ignored. The results for the discrimination setup (Eq.~(\ref{detect})) are obtained directly from Eq.~(\ref{Estatestrans}).

\subsection{$d$-dimensional $d$-photon state discrimination setup}

Our proposed setup can be easily generalized to a setup for $d$-dimensional $d$-photon entangled state discrimination. For $d$-dimensional $d$-photon entangled state discrimination, we need to generalize the three-path operation in Eq.~(\ref{unitary3}) to a $d$-path operation, which is described by a $d\times d$ discrete Fourier transform operation $\hat{U}_{d~}$, as shown in Eq.~(\ref{dDFT}):

\begin{align}\label{dDFT}
\hat{U}_{d}=\frac{1}{\sqrt{d}}\begin{pmatrix}
1 & 1 & 1 & 1 & \cdots & 1\\
1 & \chi & \chi^{2} & \chi^{3} & \cdots & \chi^{d-1}\\
1 & \chi^{2} & \chi^{4} & \chi^{6} & \cdots & \chi^{2(d-1)}\\
1 & \chi^{3} & \chi^{6} & \chi^{9} & \cdots & \chi^{3(d-1)}\\
\vdots & \vdots & \vdots & \vdots & \ddots  & \vdots\\
1 & \chi^{d-1} & \chi^{2(d-1)} & \chi^{3(d-1)} & \cdots & \chi^{(d-1)(d-1)}
\end{pmatrix}
\end{align}
where $\chi=\exp(2\pi i/d)$. The unitary operation $\hat{U}_{d}$ can be realized by means of a tritter with $d$ inputs and $d$ outputs. As an extension of the tripartite entangled qutrit state discrimination setup, the $d$-dimensional $d$-photon entangled state discrimination setup consists of a $d$-input $d$-output tritter, $d$ nondestructive photon number measurements, and $d^{2}$ on-off detectors. One of the states that the generalized setup can discriminate is very similar to the state that can be discriminated by means of the existing setup \cite{Goyal2014}, this state is given in Eq.~(\ref{dEstate}):

\begin{align}\label{dEstate}
\ket{\Phi_{0}}=\frac{1}{\sqrt{d!}}\det(\Lambda)\ket{\text{vac}}
\end{align}
where
\begin{align}
\Lambda=\begin{pmatrix}
\hat{a}^{\dagger}_{00} & -\hat{a}^{\dagger}_{01} & \cdots & (-1)^{(d-1)}\hat{a}^{\dagger}_{0(d-1)}\\
\hat{a}^{\dagger}_{10} & \hat{a}^{\dagger}_{11} & \cdots & \hat{a}^{\dagger}_{1(d-1)}\\
\vdots & \vdots & \ddots & \vdots\\
\hat{a}^{\dagger}_{(d-1)0} & \hat{a}^{\dagger}_{(d-1)1} & \cdots & \hat{a}^{\dagger}_{(d-1)(d-1)}
\end{pmatrix}
\end{align}
and $\ket{\text{vac}}$ means the vacuum state. Let us recall that in our notation, $\hat{a}^{\dagger}_{xy}$ is the photon creation operator whose orthonormal mode is $x$ and whose path label is $y$; thus, it can be rewritten as $\hat{a}^{\dagger}_{xy}\ket{\text{vac}}=\ket{x_{y}}$. If we extend the state with respect to the first row, then the state is rewritten as shown in Eq.~(\ref{dEstated}):

\begin{align}\label{dEstated}
\ket{\Phi_{0}}=&\frac{1}{\sqrt{d!}}\det(\Lambda)\ket{\text{vac}}\\
=&\frac{1}{\sqrt{d!}}\sum_{i=0}^{d-1}\hat{a}^{\dagger}_{0i}\det(\Lambda_{0i})\ket{\text{vac}} \nonumber\\=&\frac{1}{\sqrt{d}}\sum_{i=0}^{d-1}\ket{0_{i}}\ket{A_{i}}, \nonumber
\end{align}
where
\begin{align}
\ket{A_{i}}=\frac{1}{\sqrt{(d-1)!}}\det(\Lambda_{0i})\ket{\text{vac}}
\end{align}
and $\Lambda_{0i}$ is the $(d~-~1)\times(d~-~1)$ submatrix obtained by omitting the $(i~+~1)$th column and first row of $\Lambda$. Similar to the case of tripartite entangled qutrit state discrimination, all states that the generalized setup can discriminate are orthogonal to $\ket{\Phi_{0}}$ with a phase factor $\chi$; these states can be described as given in Eq.~(\ref{dEstates}):

\begin{align}\label{dEstates}
\ket{\Phi_{i}}=\frac{1}{\sqrt{d}}\sum_{j=0}^{d-1}\chi^{ij}\ket{0_{j}}\ket{A_{j}},
\end{align}
where $i\in\{0,1,2,3,...,d-1\}$. These $d$ states can be discriminated from the combinations of clicked detectors in the generalized setup after post-selection using nondestructive photon number measurements.

\section{Acknowledgements}
We acknowledge the financial support from the National Security Research Institute (NSRI) and the R\&D Convergence Program of the National Research Council of Science and Technology (NST)  of Republic of Korea (Grant No. CAP-15-08-KRISS). W. Son acknowledges the support provided by KIAS through the Open KIAS Program.

\section{Author contributions statement}
Y.J. designed and analysed the protocols. K.B. contributed to the analysis. W.S. supervised the whole project. All authors reviewed the manuscript.

\section{Additional information}
\textbf{Competing financial interests}:
Y. Jo was funded by the Agency for Defense Development (ADD). K. Bae and W. Son were funded by The National Security Research Institute (NSRI). W. Son received a salary from Sogang University. W. Son also acknowledges the University of Oxford and KIAS for their visitorship programme.
\end{document}